\documentclass[aps,prc,tightenlines,floatfix,showpacs,twocolumn]{revtex4-1}

\usepackage{graphics}
\usepackage{bm}
\usepackage{amsmath}
\usepackage{amssymb}
\usepackage{supertabular}
\usepackage{array} 

\newcommand{\sfrac}[2]{\textstyle\frac{#1}{#2}}
\newcommand{\pdif}[2]
{\ensuremath{\frac{\partial{#1}}{\partial{#2}}}}

\begin{document}

\title{Coupling to two target-state bands in the study of the $n+^{22}$Ne 
system at low energy.}

  \author{P. R. Fraser$^{(1,2,3)}$}
  \email{prfraser@unimelb.edu.au}
  \author{L. Canton$^{(1)}$}
  \author{K. Amos$^{(2,4)}$}
  \author{\mbox{S. Karataglidis$^{(2,4)}$}}
  \author{J. P. Svenne$^{(5)}$}
  \author{D. van der Knijff$^{(2)}$}

\affiliation{$^{(1)}$ Istituto  Nazionale  di  Fisica  Nucleare,
Sezione di Padova, Padova I-35131, Italia}
\affiliation{$^{(2)}$ School  of Physics,  University of  Melbourne,
Victoria 3010, Australia}
\affiliation{$^{(3)}$ Department of Imaging and Applied Physics,
Curtin University, Bentley, Western Australia 6102, Australia}
\affiliation{$^{(4)}$ Department of Physics, University of
Johannesburg, P.O. Box 524 Auckland Park, 2006, South Africa}
\affiliation{$^{(5)}$ Department  of  Physics  and Astronomy,
University of Manitoba, and Winnipeg Institute for Theoretical
Physics, Winnipeg, Manitoba, Canada R3T 2N2}

\date{\today}

\begin{abstract}
One theoretical method for studying nuclear scattering and resonances
is via the multi-channel algebraic scattering (MCAS)
formalism. Studies to date with this method have used a simple
collective-rotor prescription to model target states with which a
nucleon couples. While generally these target states all belong to the
same rotational band, for certain systems it is necessary to include
coupling to states outside of that main band. Here, we extend MCAS to
allow coupling of different strengths between such states and the
rotor band. This is an essential consideration in studying the example
examined herein, the scattering of neutrons from $^{22}$Ne.
\end{abstract}
 
\maketitle

\section{Introduction}

As an example of the effects of considering states that weakly couple
to those within a collective band, we examine neutron scattering from
$^{22}$Ne. Besides the study of the mass-23 isobars being of interest
inherently, the formation of these nuclei by radiative capture is of
great astrophysical interest~\cite{Ti03,Ti05}. For example, it is
important to understand the processes leading to their presence in
white dwarf stars, as $^{23}$Ne and $^{23}$Na form an Urca
pair~\cite{Ri99}, emitting neutrinos and delaying a supernova
explosion. Such type Ia supernovae have properties which are thought
usable to measure the extent and expansion of the universe.

Another topical problem associated with these isobars is the so-called
$^{22}$Na puzzle of ONe white dwarf novae~\cite{Ba11,Al10,Ji13}, where
the abundance of $^{22}$Na predicted by existing stellar models is not
found, indicating there is yet more to learn about how the
distribution of elements in the universe occurred.  Two reactions
possibly pertinent to this loss of abundance are
$^{22}$Mg$(p,\gamma)^{23}$Al and $^{22}$Na$(p,\gamma)^{23}$Mg. MCAS is
well suited to modelling the $^{22}$Mg$(p,p)^{22}$Mg reaction (the
mirror is the system studied here) due to the low density of
low-energy states in $^{23}$Al and the low scattering threshold.  This
is a necessary first step in analysing the resonant capture
$^{22}$Mg$(p,\gamma)^{23}$Al using the formalism of
Ref.~\cite{Ca08}. Development of the MCAS project to obtain capture
cross-section values is in progress.

This study is also a prelude to that of the $p+^{22}$Ne system, since
that scattering and the associated capture cross section is
important in astrophysics. It is part of the hydrogen-burning NeNa
cycle which may occur in second-generation stars. Speculated leakage
from the CNO cycles into the NeNa cycle is linked to the problem of
anti-correlations having been observed between sodium and oxygen when
stars ascend the red giant branch of the Hertzsprung-Russell diagram,
despite current stellar models predicting that the surface abundance
of elements should not change.  The rate of the
$^{22}$Ne$(p,\gamma)^{23}$Na reaction depends on the strengths of
several resonances which have never been observed experimentally.
Being afflicted by extremely large uncertainties~\cite{Il02},
theoretical treatment of both resonant and direct capture will be
desirable to complement experimental investigations performed at
LUNA~\cite{Ca12} at $E < 400$ keV. Indeed, they are topics of future
work.

The multi-channel algebraic scattering (MCAS) formalism~\cite{Am03} is
one with which scattering observables and spectra for quantum systems
can be evaluated. To date it has been used for nuclear processes. In
the MCAS method, solutions of coupled-channel Lippmann-Schwinger
equations are found in momentum space using finite-rank expansions of
an input matrix of nucleon-nucleus interactions. A set of sturmian
functions is used as the expansion basis. The MCAS method is able to
locate all compound-system resonance centroids and widths, regardless
of how narrow, and has the ability to determine subthreshold bound
states by using negative energies. Further, use of orthogonalizing
pseudo-potentials (OPP) in generating sturmians, ensures that the
Pauli principle is not violated~\cite{Ca05,Am13}, even with a
collective model formulation of nucleon-nucleus
interactions. Otherwise, some compound nucleus wave functions possess
spurious components~\cite{Am05}.

Having the purview of low-energy scattering, with a range of a few
MeV, MCAS usually deals with target nuclei in which only one mode of
collective behaviour is exhibited. To date, the Tamura~\cite{Ta65}
collective model with rotational character has been used to determine
a coupled-channel interaction with nucleons. Thus, target states
selected for coupling should be from within the principal rotation
band of nuclei which exhibit such behavior, and the $\beta_L$ (which
determines the coupling strengths) is the same coupling amongst all
states.  Often, these calculations reproduce scattering observables
very well, and even possess predictive power~\cite{Ca05,Fr08a,Am12}.
At times, however, it is necessary to include coupling to states
outside of this band, for example where experiment has shown
$\gamma$-decays to states within a clearly-defined collective
band. Here, we extend MCAS to allow coupling of different strengths
between such states and those in the rotor band.

Section~\ref{sec-formalism} shows details of the development of MCAS
rotor potentials for $NA$ scattering, adding two facets to that
previously published~\cite{Am03}: a generalisation to allow more than
one multipole deformation, and an extension to allow a second band for
a given deformation. As $n+^{22}$Ne is the system selected to
illustrate these developments, Section~\ref{sec-shell} shows results
of a no-core $(0+2)\hbar\omega$ shell model calculation for $^{22}$Ne,
allowing insight into structure of orbit occupancy of the target
states. Section~\ref{sec-Ne23} shows a method of identifying the ratio
of $\beta_2$ values linking second-band states to first-band states,
examines the effect of having different $\beta_2$ bands on the
calculated spectrum, and determines which channels are important in
describing the $n+^{22}$Ne system. Results are shown for both spectra
and elastic cross section. Finally, in Section~\ref{sec-conclusion},
conclusions are drawn.

\section{A tale of two $\beta_L$s}
\label{sec-formalism}

To illustrate the manner in which different coupling strengths between
channels stemming from different target states are considered, the
development of the rotational-type coupled-channel $NA$-scattering
potential is summarized. How these potentials are treated with the
MCAS solution of the coupled-channel Lippmann-Schwinger equations is
covered in detail in Ref.~\cite{Am03}.

With channels defined by
\begin{equation}
c = \left[(l \sfrac{1}{2})j I; J M \right] \ ,
\label{channelsROT}
\end{equation}
$(l \sfrac{1}{2})j$ are the orbital, intrinsic spin, and total angular
momentum of relative motion of the projectile on the target, $I$ is the
total angular momentum of the target state involved, and $J,M$ are the
angular momentum quantum numbers of the compound system.
Then, we define a $NA$ coupled-channels potential matrix by:
\begin{multline}
V_{cc'}(r) =
f(r) \biggl\{ V_0 \delta_{cc'} + V_{ll} [ {\bf {\ell \cdot 
\ell}} ]_{cc'}\\
+ V_{ss} [ {\bf {s \cdot I}} ]_{cc'} \biggr\} +  g(r) V_{ls}
[{\bf {\ell \cdot s}}]_{cc'}\ ,
\label{www}
\end{multline}
in which local form factors have been assumed, and parameters of the
potential governing central ($V_0$), orbit-orbit ($V_{ll}$), spin-spin
($V_{ss}$), and spin-orbit ($V_{\ell s}$) components. (Note that,
being a parameter of the model, $V_{ls}$ contains the constant
$2\lambda_\pi^2$, where $\lambda_\pi$ is related to the inverse of the
pion Compton wave length.) We identify the functions $f(r)$ and $g(r)$
with deformed Woods-Saxon form factors:
\begin{equation}
f(r) = \left[1 + e^{\left( {{r-R}\over a} \right)} \right]^{-1}
\hspace*{0.3cm} ; \hspace*{0.3cm} g(r) = \frac{1}{r} \frac{df(r)}{dr} \: .
\end{equation}

To introduce a rotor character for this general nucleon-nucleus
interaction potential, let us first consider that the quantum radius
of a rigid drop of nuclear matter, with axial, permanent deformation
from the spherical, is represented by the expansion
\begin{align}
R(\theta, \phi)  &= R_0 \left[ 1 + 
\sum_{L (\ge 2)} \sqrt{\frac {4 \pi}{2L+1}}\  \beta_L \
[{\bf Y}_L (\hat r) {\bf \cdot} {\bf Y}_L(\hat\Upsilon)] \right]\nonumber\\ 
&= R_0 \left[1 + \epsilon \right]\ ,
\label{SurfaceRotor}
\end{align}
where $(\hat{r}) = (\theta,\phi)$ designates internal target
coordinates. $\hat\Upsilon$ are Euler angles specifying the
transformation from body-fixed to space-fixed frame co-ordinates.

Expanding $f(r)$ in Eq.~(\ref{www}) to second order in $\epsilon$ gives
\begin{equation}
f(r) = f_0(r) + \epsilon \left(\pdif{f(r)}{\epsilon} \right)_0 
+ \frac{1}{2} \epsilon^2 \left(\pdif{^2f(r)}{\epsilon^2} \right)_0\ . 
\end{equation}
We wish to convert these derivatives to being in terms of $r$. If we
demand that $f(r) = f(r-R(\theta, \phi))$, that for every $r$ there is
an accompanying subtraction of $R$, we use the following:
\begin{equation}
\pdif{f(r)}{\epsilon}
= \pdif{f(r-R)}{R} \pdif{R}{\epsilon} = - R_0 \ \pdif{f(r-R)}{r}\ .  
\label{r-R}
\end{equation}
Thus,
\begin{multline}
f(r) =  f_0(r) \\ - R_0 \sum_{L (\ge 2)} \sqrt{\frac {4 \pi}{2L+1}}\  \beta_L \
[{\bf Y}_L (\hat r) {\bf \cdot} {\bf Y}_L(\hat \Upsilon)]\ \frac{df_0(r)}{dr}
\\
+ \frac{1}{2} R_0^2 \left[ \sum_{L (\ge 2)} \sqrt{\frac {4 \pi}{2L+1}}\ 
\beta_L \ [{\bf Y}_L (\hat r) {\bf \cdot} {\bf Y}_L(\hat \Upsilon)] \right]^2 
\frac{d^2f_0(r)}{dr^2}\ .
\label{fff}
\end{multline}
 
Keeping $L$ general, and not assuming $L = L'$, i.e., that only
one $\beta$ of deformation is considered, we obtain
\begin{eqnarray}
\epsilon^2 &=& \left[\sum_{L (\ge 2)} \sqrt{\frac {4 \pi}{2L+1}}\  \beta_L \
[{\bf Y}_L (\hat r) {\bf \cdot} {\bf Y}_L(\hat \Upsilon)]\ \right]^2
\nonumber\\
           &=& \sum_{L,L' (\ge 2)}
\frac{4 \pi \beta_L \beta_{L'}}{\sqrt{(2L+1)(2L'+1)}}\nonumber\\
&&\hspace{1cm}\times[{\bf Y}_L (\hat r) {\bf \cdot} {\bf Y}_L(\hat \Upsilon)]\
[{\bf Y}_{L'} (\hat r) {\bf \cdot} {\bf Y}_{L'}(\hat \Upsilon)].
\label{epssq}
\end{eqnarray}
Using a property of tensor products~\cite{Va88}, we can express
\begin{align}
&[{\bf Y}_L (\hat r) {\bf \cdot} {\bf Y}_L(\hat \Upsilon)]\
[{\bf Y}_{L'} (\hat r) {\bf \cdot} {\bf Y}_{L'}(\hat \Upsilon)]
\nonumber\\
&=\ \frac{(2L+1)(2L'+1)}{4 \pi}\times \nonumber\\
&\hspace{1cm} \sum_{\ell} \frac{1}{2 \ell +1}\
|\langle L 0 L' 0|\ell 0 \rangle|^2\ 
[{\bf Y}_{\ell} (\hat r) {\bf \cdot} {\bf Y}_{\ell}(\hat \Upsilon)]\ ,
\label{yprod}
\end{align}
where $\ell$ runs from $|L-L'|$ to $L+L'$, with the condition that $L
+ \ell + L'$ is even. Thus,
\begin{multline}
f(r) =  f_0(r) \\ - R_0 \sum_{L (\ge 2)} \sqrt{\frac {4 \pi}{2L+1}}\  \beta_L \
[{\bf Y}_L (\hat r) {\bf \cdot} {\bf Y}_L(\hat \Upsilon)]\ \frac{df_0(r)}{dr}
\\
+ \frac{1}{2} R_0^2 
\sum_{L,L' (\ge 2)} \beta_L \beta_{L'} {\sqrt{(2L+1)(2L'+1)}}
\\
\hspace{0.6cm} \times \sum_{\ell} \frac{1}{2 \ell +1}\
|\langle L 0 L' 0|\ell 0 \rangle|^2\ 
[{\bf Y}_{\ell} {\bf \cdot} {\bf Y}_{\ell}]
\frac{d^2f_0(r)}{dr^2}\ .
\label{fff2}
\end{multline}
A similar equation applies for the expansion of $g(r)$ in terms of the
deformation $\epsilon$.

As the full potential is now rather detailed, we consider it in terms of
its zeroth, first and second order expansion components:
\begin{align}
V_{cc'}(r) =& V_{cc'}^{(0)}(r) + V_{cc'}^{(1)}(r) + V_{cc'}^{(2)}(r)
\nonumber\\
=& \left\{v^{(0)}(r)\right\}_{cc'}\nonumber\\  
&+\  \left\{v^{(1)}(r)\ 
\sum_{L (\ge 2)} \beta_L\ \sqrt{\frac{4\pi}{2L+1}}\  
\left[ {\bf Y}_L {\bf \cdot} {\bf Y}_L \right] 
\right\}_{cc'} 
\nonumber\\
&+\; \left\{v^{(2)}(r)\ 
\sum_{L,L' (\ge 2)} \beta_L \beta_{L'} {\sqrt{(2L+1)(2L'+1)}} \right.\nonumber\\
&\hspace{0.6cm} \left. \times \sum_{\ell} \frac{1}{2 \ell +1}\
|\langle L 0 L' 0|\ell 0 \rangle|^2\ 
[{\bf Y}_{\ell} {\bf \cdot} {\bf Y}_{\ell}]
\right\}_{cc'} .
\label{Mcaspot-generalL}
\end{align}

This is a short-hand notation; as we focus on the $\beta_L$, the
functions $v^{(0)}(r)$, $v^{(1)}(r)$ and $v^{(2)}(r)$ are employed to
subsume all terms independent of $L$, concerning the derivatives of
the Woods-Saxon form factors and potential variables. For
completeness, these are shown in full in the Appendix, where the
interplay of spin-angular operators and multipole deformations are
taken into account.

The above development is similar to that of Ref.~\cite{Am03}, but is
generalised to consider more than one multipole deformation; that is,
cases where $L \ne L'$. (N.B. This development has been used
previously~\cite{Am12}, but heretofore has not been presented in detail.)

Next, we consider cases where there exist states which are outside the
main rotational band, but which are known to couple to states in the
rotor band. To describe this weaker coupling, it is necessary to
include, for a given $L$, an additional value of $\beta_L$, which we
denote here as $\overline{\beta_L}$. This can be done with a scaling,
\textit{viz.}
\begin{equation}
\overline{\beta_L} = s_L \beta_L \: ,
\label{sL}
\end{equation}
whereby Eq.~(\ref{Mcaspot-generalL}) becomes
\begin{align}
&V_{cc'}(r) = \left\{v^{(0)}(r)\right\}_{cc'}\nonumber\\  
&\:\:+\  \left\{v^{(1)}(r)\ 
\sum_{L (\ge 2)} s_L \beta_L\ \sqrt{\frac{4\pi}{2L+1}}\  
\left[ {\bf Y}_L {\bf \cdot} {\bf Y}_L \right] 
\right\}_{cc'} 
\nonumber\\
&\:\:+\; \left\{v^{(2)}(r)\ 
\sum_{L,L' (\ge 2)} s_L \beta_L s_{L'}\beta_{L'} {\sqrt{(2L+1)(2L'+1)}}
\right.\nonumber\\
&\hspace{0.6cm} \left. \times \sum_{\ell} \frac{1}{2 \ell +1}\
|\langle L 0 L' 0|\ell 0 \rangle|^2\ 
[{\bf Y}_{\ell} {\bf \cdot} {\bf Y}_{\ell}]
\right\}_{cc'} ,
\label{Mcaspot-generalLmod}
\end{align}
\begin{equation*}
\text{where}
  \begin{cases}
   s_L = 1     &\text{if }\displaystyle I^\pi_i\bigr\rvert_c \text{ and } 
                 I^\pi_j\bigr\rvert_{c'} \in \text{main band} \\
   0 < \left|s_L\right| < 1 &\text{if }\displaystyle I^\pi_i\bigr\rvert_c 
                 \text{ and/or }
                 I^\pi_j\bigr\rvert_{c'} \notin \text{main band}, \\
  \end{cases}
\end{equation*}
$I^\pi_i$ being the $i^{th}$ $I^\pi$ target state, following the
convention of the channel definition in Eq.~(\ref{www}).  These
changes of band correspond to a shape transition. In future works we
intend to refine this coupling scheme such that it takes into account
differences between reorientation within a given band and transitions
between different bands.

\section{A shell model for states in ${}^{22}{\rm Ne}$}
\label{sec-shell}

Before considering scattering of neutrons from $^{22}$Ne as an example
of the expansion of MCAS considered here, it is instructive to
consider what can be gleaned about that target nucleus from
shell-model studies of adequate complexity as has been used in
Ref.~\cite{Ka95,Ka96,Am13a}.

We first sought results from a no-core $(0+2)\hbar \omega$ shell-model
for ${}^{22}$Ne.  The OXBASH program~\cite{Ox86} with the WBT
interactions~\cite{Wa92} was used. The single nucleon space chosen
encompassed the 15 orbits in shells from the $1s_{\frac{1}{2}}$
through the $1g_{\frac{9}{2}}$-$3s_{\frac{1}{2}}$, Those evaluations
involved such large dimensioned matrices that only vectors and energy
values of the positive parity $J\le 2$ states in the low-energy
excitation spectrum of ${}^{22}$Ne could be found.  Allowing all 22
nucleons to be active is beyond the capacity of the standard
OXBASH program we have used to find higher spin states and, as the
associated vectors are very large, we have been unable as yet to
extract many properties of those states. The results given then are
preliminary to a planned fuller study which will include more nuclei
in the mass region and made using a larger shell model program.

We have also made calculations within a reduced ($0\hbar\omega$)
space, the $1d-2s$ shell for ${}^{22}$Ne, to give some indication of
the major shell transition strengths between the $2^+_1$ and $2^+_2$
and the ground states as those three are of special interest in the MCAS
studies.

First consider the (preliminary) results found using the large-space
shell model.  The evaluated excitation energies for the low-lying
$0^+$ and $2^+$ states are in good agreement with data, as is evident
in the listing in Table~\ref{sm-levels}.
\begin{table}[h]
\begin{ruledtabular}
\caption{\label{sm-levels}
The low lying $0^+$ and $2^+$ state energy levels in ${}^{22}$Ne
compared with values determined using the large space shell model
calculation described in the text. Energies are in MeV and component types
are in percent.}
\begin{tabular}{ccccc}
state & Exp. & shell model & $0\hbar\omega$ & $2\hbar\omega$ \\
\hline
$0^+_{rm g.s.}$ & 0.000 & 0.000 & 67.62 & 32.38 \\
$2^+_1$         & 1.275 & 1.336 & 67.48 & 32.51 \\
$2^+_2$         & 3.358 & 4.244 & 67.15 & 32.85 \\
$2^+_3$         & 4.456 & 4.507 & 66.89 & 33.11 \\
$2^+_4$         & 5.363 & 5.579 & 66.57 & 33.43 \\
$2^+_5$         & 6.120 & 6.185 & 67.00 & 33.00 \\
$0^+_2$         & 6.234 & 5.803 & 66.72 & 33.28 \\
$0^+_3$         &       & 6.428 & 66.39 & 33.61 \\
\end{tabular}
\end{ruledtabular}
\end{table}
Also shown in the table are the percentage admixture of $0\hbar
\omega$ and $2\hbar \omega$ components in each state description. All
states are considerably mixed with, characteristically, 33\% of
$2\hbar\omega$ component.

Further, all states are specified by numerous partitions of the
nucleons within the orbits. Those contributing the largest percentages
(greater than 5\%) are listed in Table~\ref{gs-part}.  These dominant
partitions have the $1s-1p$ shells completely full (occupancies 4, 8,
and 4) and those for the remaining 6 nucleons (2 protons and 4
neutrons) are listed according to the shell indicated.
\begin{table}[h]
\caption{\label{gs-part} Dominant partition (total nucleons) percentages
(values $\ge 5$\%) in the
  shell model ground, $0^+_2$, $2_1^+$, and $2^+_2$ states of ${}^{22}$Ne.} 
\begin{ruledtabular}
\begin{tabular}{ccccc|cc}
 & & & $0^+_{\rm g.s.}$ & $0^+_2$ (5.803) & $2^+_1$ (1.336) & $2^+_2$ (4.244) \\
\hline
$1d_{\frac{5}{2}}$ & $1d_{\frac{3}{2}}$ & $2s_{\frac{1}{2}}$ & Percent & Percent &
 Percent & Percent \\
\hline
4 &  2 & 0 & \ 7.82 & & & \\
4 &  1 & 1 & & \ 5.59 & \ 7.84 & \ 8.45 \\
5 &  1 & 0 & \ 5.64  & & \ 4.57 & \ 7.11 \\
3 &  1 & 2 & & \ 9.21 & & \\
5 &  0 &  1 & \ 6.67 & \ 8.13 & 15.65  &  12.08 \\
4 &  0 &  2 & 12.47 & 21.59 & \ 6.43 & \ 6.88 \\
6 &  0 &  0 & 20.05 & \ 5.00 & 14.16  &  14.12 
\end{tabular}
\end{ruledtabular}
\end{table}
This shell model gave another 11 partitions for these states all
having percentages of between 1 and 5\%; 4 having a reduced occupancy
in the $1p$-orbits (offset by some in the $2p$-orbits) and 2 more with
occupancy in the $2d$-orbit.  The $1d$-$2s$ shell is the most
important in this description of these states, with components are
spread over all three orbits of that shell, but the $\sim 33$\%
involving the other shells is needed to find the best result for the
spectrum.  There are also numerous other entries having smaller
($<$1\%) amplitudes.

Including $2\hbar\omega$ components in shell model descriptions of
nuclear states of several light-mass nuclei has lead to predictions of
transition rates enhanced on those found limiting the structure
evaluations to $0\hbar\omega$. Often the latter models require a
significant polarisation charge to give a match to measured B(E2)
rates and electron scattering form factors, for example, while some
studies using larger space structures do not~\cite{Ka95,Ka96}.
Nevertheless we next present results obtained using a $0\hbar\omega$
model (only the $1d-2s$-shell active with the USD interaction of Brown
and Wildenthal~\cite{Br88}) to illustrate that the $2^+_1$ and $2_2^+$
states should both have non-negligible transition strength to the
ground, though coupling to the $2^+_1$ is dominant.  The one body
density matrices that link the $2^+_1$ and the $2^+_2$ states to the
ground are given in Table~\ref{sm-obd}.
\begin{table}[h]
\begin{ruledtabular}
\caption{ \label{sm-obd}
The shell model one-body-density matrix values in the $2s$-$1d$ shell
linking the $2^+_1$ and $2^+_2$ states to the ground.}
\begin{tabular}{cccc}
 $j_1$ & $j_2$ & $2^+_1$ (1.366) & $2^+_2$ (4.244) \\
\hline
$1d_{\frac{5}{2}}$ & $1d_{\frac{5}{2}}$ & $-$0.9582 & $-$0.1153 \\
$1d_{\frac{5}{2}}$ & $1d_{\frac{3}{2}}$ & $-$0.3261 & \ 0.1030 \\
$1d_{\frac{5}{2}}$ & $2s_{\frac{1}{2}}$ & $-$0.5945 & $-$0.1140 \\
$1d_{\frac{3}{2}}$ & $1d_{\frac{5}{2}}$ & \ 0.3341 & \ 0.0503 \\
$1d_{\frac{3}{2}}$ & $1d_{\frac{3}{2}}$ & $-$0.0886 & $-$0.1142 \\
$1d_{\frac{3}{2}}$ & $2s_{\frac{1}{2}}$ & \ 0.2472 & \ 0.1433 \\
$2s_{\frac{1}{2}}$ & $1d_{\frac{5}{2}}$ & $-$0.6954 & $-$0.0830\\
$2s_{\frac{1}{2}}$ & $1d_{\frac{3}{2}}$ & $-$0.2073 & \ 0.0118\\
\end{tabular}
\end{ruledtabular}
\end{table}
These quantities are defined by the (doubly reduced) matrices
for $\Delta T = 0$, namely
\begin{equation}
S_{j_1 j_2 I=2} = \left\langle 0^+_{\rm g.s.} \left|\left|\left|
\left[ a_{j_2}^\dagger \otimes {\tilde a}_{j_1} \right]^{(I=2)}
\right|\right|\right| 2^+_{(1,2)} \right\rangle\ .
\end{equation}
It is clear that, from these shell model results, ground state
coupling favours the $2^+_1$ state, but there is some non-negligible
strength to the $2^+_2$ state; of between $10$ and $30$\% for most
terms.

\section{Initial MCAS evaluation of the ${\rm n}$+${\rm ^{22}Ne}$ system}
\label{sec-Ne23}

The low-lying spectrum of $^{22}$Ne consists of ground state of $J^\pi
= 0^+$, a $2^+$ state at 1.274 MeV, and a $4^+$ state at 3.357
MeV. Directly above this comes a $2^+$ state at 4.456 MeV which decays
by E2 transition to the ground state~\cite{Fi05}. 

\begin{figure}[htp]
\begin{center}
\scalebox{0.7}{\includegraphics*{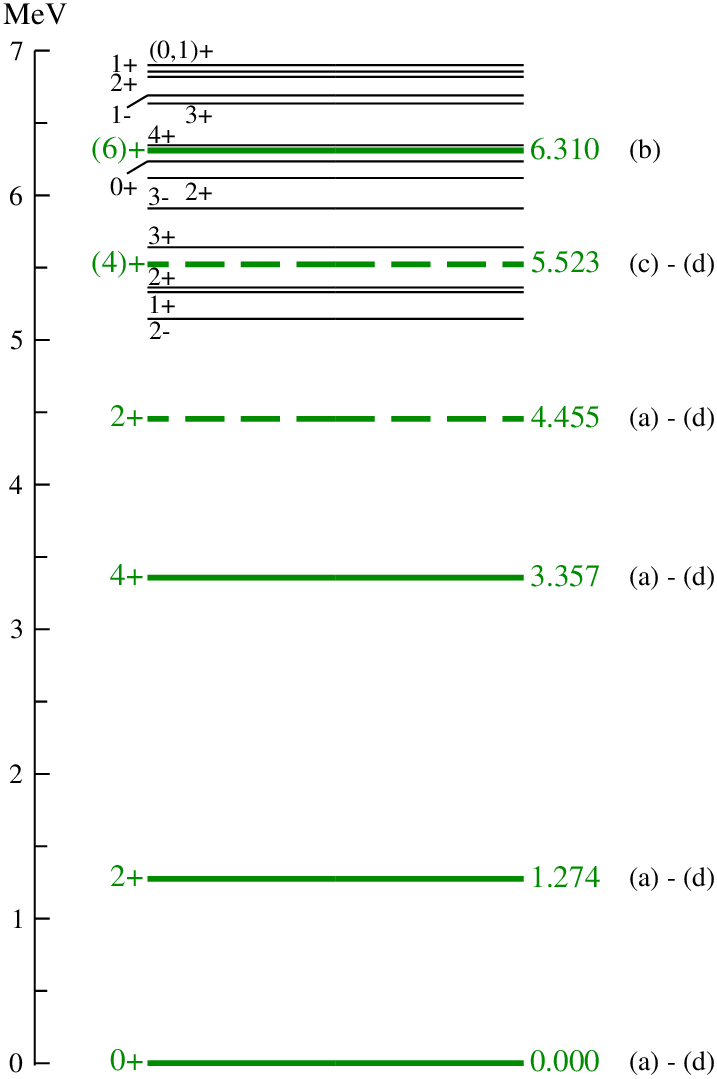}}
\end{center}
\caption{ \label{Ne22-spec}(Color online.) The low-energy experimental
  $^{22}$Ne spectrum~\cite{Fi05}. Thick, solid lines denote states of
  the ground-state band, thick dashed lines denote the other states
  used in this paper. Letters correspond to usage in specific
  calculations in Figs.~\ref{Ne23-spec}, \ref{Ne23-spec-6+} and
  \ref{Ne23-spec-4+}.}
\end{figure}

The $0^+$, $2^+$, and $4^+$ states, along with a state at 6.31 MeV
designated $(6)^+$ in the literature~\cite{Fi05}, we characterise as a
rotor behaviour. The actual spectrum of $^{22}$Ne to 7 MeV excitation
is shown in Fig.~\ref{Ne22-spec}. The rotor-like spacing of the
principal band (shown in thick, solid lines) is evident.

\subsection{$\beta_2$ values for the two $2^+$ states in ${}^{22}$Ne.}
\label{beta1}

Given that the $2^+_2$ state (shown as a dashed line in
Fig.~\ref{Ne22-spec}) decays to the ground state but is not within the
sequence of the first rotation-like band, we can expect that it
exhibits some other degree of rotor character. Thus, we assign a
different $\beta_2$ value (denoted $\overline{\beta_2}$ to distinguish
it from that used for the main band) to link this state with those
taken to be the principal rotor band.

The half lives of states (ground state $\gamma$ decay)
relate to the transition probabilities  (for $E2$ multipolarity)
via
\begin{equation}
\tau_{\frac{1}{2}} = 
\frac{\ln(2)}{{\cal W}_{(E2)}(E_\gamma)}
= \frac{0.693}{{\cal W}_{(E2)}(E_\gamma)} \,:
\end{equation}
and transition probabilities link to $B(E2)$ values via
\begin{equation}
{\cal W}_{(E2)}(E_\gamma) = 1.23 \times 10^9\ \left(E_\gamma \right)^5
\ B(E2) ,
\end{equation}
where $E_\gamma$ is the photon energy.

To first order, and without consideration of band quantum numbers, a
collective (rotational) model gives $B(E2)$ that are proportional to
$\beta_2^2$.

For ${}^{22}$Ne, the two low excitation $2^+$ states (at 1.275 and
4.456 MeV) both decay by $\gamma$-emission via $E2$ transitions to the
ground state with half lives of 3.63 ps and by 37 fs, respectively.
Thus the relevant transition probabilities are
\begin{align}
{\cal W}_{(E2)}(1.275) &= 
\frac{0.693}{3.63} 10^{12}
\nonumber\\
&= 1.23 \ 10^9\ (1.275)^5 \ B(E2, 1.275)
\nonumber\\
{\cal W}_{(E2)}(4.456) &=
\frac{0.693}{37} 10^{15}
\nonumber\\ 
&= 1.23 \ 10^9\ (4.456)^5 \ B(E2, 4.456) ,
\end{align}
from which $B(E2, 1.275) = 46.06$ and $B(E2, 4.456) = 8.67$ (units are
$e^2$ fm) and their ratio is 0.188. Then assuming that the $B(E2)$
scale as $\beta_2^2$, the deformation length for the 4.456 MeV decay
would be ${\sim}0.43$ times that for the 1.275 MeV decay, i.e.,
$\overline{\beta_2} = 0.43\beta_2$.

\subsection{Results using the lowest four target states}
\label{4-state}

For a scattering nucleon impinging upon a partially-filled shell of a
target nucleus, the Pauli principle does not necessarily imply a
binary rule - that the shell is completely open or completely
blocked~\cite{Sc78,Sc80,Ba81,Sc82,La86,Fi04,La07}. We label as Pauli
hindrance the intermediate situation, where the present nucleons do
not completely, but only partially block additional
nucleons. Considering such an interpretation, the dominant
configurations of the shell model descriptions of the states in
${}^{22}$Ne, shown in Table~\ref{gs-part}, that are of particular
interest in MCAS calculations prescribe full blocking of the
$1s\sfrac{1}{2}$, $1p\sfrac{3}{2}$, and $1p\sfrac{1}{2}$ orbits, while
suggesting Pauli hindrance for the
$1d{\sfrac{5}{2}},1d{\sfrac{3}{2}},2s{\sfrac{1}{2}}$ orbits, meaning
an orthogonalising pseudo-potential (OPP) that creates only a partial
blocking of those $d-s$ orbits. Those shell model functions further
suggest that all higher subshells have essentially no blocking.

For simplicity, however, in this study we have considered purely
allowed or purely blocked states. Consequently, parameters used to
scale these OPP, denoted as $\lambda^{(OPP)}$ and in units of MeV, are
assigned a value of 10$^6$~MeV (which is adequate to remove all
influence of blocked states) for the orbitals $1s\sfrac{1}{2}$,
$1p\sfrac{3}{2}$, and $1p\sfrac{1}{2}$, and 0 MeV for
$1d\sfrac{5}{2}$, as shown in Table~\ref{params}. Full details of the
Pauli principle in MCAS, including the blocking strengths of the OPP
method, are given in Refs.~\cite{Ca05,Am13}.

The parameter set used to define the scattering potential is shown in
Table~\ref{params}. As with Ref.~\cite{Sw72}, which studied analysing
powers from $^{22}$Ne$(p,p)^{22}$Ne, it was found that a small
$\beta_4$ deformation of $^{22}$Ne was required in MCAS. However, the
best-fit MCAS deformations differ from those of Ref.~\cite{Sw72}
(being $\beta_2 = 0.47$ and $\beta_4 = 0.05$), which is understandable
given the differences between the models; where the MCAS potential
includes $V_0$, $V_{l l}$, $V_{l s}$ and $V_{ss}$ terms and the same
radius and diffusivity for $V_{l s}$ as the other terms,
Ref.~\cite{Sw72} uses $V_0$ and $V_{l s}$ alone (with a different
prescription for the latter), but with different radius and
diffusivity for $V_0$ and $V_{l s}$. Despite this, both values are of
the same order of magnitude in each paper. We note, however, that a
value of $\beta_2 = 0.562$ was proposed in Ref.~\cite{Ra01}, based
upon an adopted value of the reduced $B(E2)\hspace{-1.4mm}\uparrow$
from the $0^+$ ground state to the $2^+$ first excited state. While
two values of $\beta_2$ are used in this work as $E2$ transitions are
observed between rotor and non-rotor-like states in $^{22}$Ne, the
same value of $\beta_4$ is used in all instances throughout the paper.
In Ref.~\cite{Ba11}, studies of the mirror system
$^{22}$Mg$(p,p)^{22}$Mg were made using Glauber-type
calculations~\cite{Sa04}. In these, a Woods-Saxon potential was used
(as it is in MCAS for nuclear and Coulomb potentials) for the proton
binding potential, with standard nucleus radius (similar to ours) and
diffuseness $a = 0.60$ fm.

\begin{table}\centering
\caption{\label{params} 
Parameter values defining the $n+^{22}$Ne interaction. $\lambda^{(OPP)}$
are blocking strengths of occupied shells, in MeV.}
\begin{supertabular}{>{\centering}p{28mm} p{27mm}<{\centering} 
p{27mm}<{\centering}}
\hline
\hline
 & Odd parity & Even parity\\
\hline
$V_0$ (MeV) & -65.20  & -51.30\\
$V_{l l}$ (MeV)       &  -1.01  &  -0.30\\
$V_{l s}$ (MeV)       &   7.00  &   7.00\\
$V_{ss}$ (MeV)        &  -0.20  &  -1.45\\
\end{supertabular}

\begin{supertabular}{>{\centering}p{14mm} p{18mm}<{\centering} 
>{\centering}p{17mm} p{14mm}<{\centering} p{17mm}<{\centering} }
\hline
\hline\\[-1.9ex]
$R_0$ & $a$ & $\beta_2$ & $\overline{\beta_2}$ & $\beta_4$\\
3.1 fm & 0.75 fm & 0.22 & 0.1034 & -0.08\\
\end{supertabular}
\begin{supertabular}{>{\centering}p{18mm} p{16mm}<{\centering} 
p{16mm}<{\centering} p{16mm}<{\centering} p{14mm}<{\centering}}
\hline
\hline
                        &$1s_{1/2}$ &$1p_{3/2}$ &$1p_{1/2}$&$1d_{5/2}$\\
\hline\\[-1.9ex]
 $0^+_1 \; \lambda^{(OPP)}$ & 10$^6$ & 10$^6$ & 10$^6$ & 0.0\\
 $2^+_1 \; \lambda^{(OPP)}$ & 10$^6$ & 10$^6$ & 10$^6$ & 0.0\\
 $4^+_1 \; \lambda^{(OPP)}$ & 10$^6$ & 10$^6$ & 10$^6$ & 0.0\\
 $2^+_2 \; \lambda^{(OPP)}$ & 10$^6$ & 10$^6$ & 10$^6$ & 0.0\\
\hline
\hline\\[-1.9ex]
\end{supertabular}
$\overline{\beta_2}$ for linking $2^+_2$ to other states; 43\% of 0.22. See
Section~\ref{beta1}.
\end{table}

\begin{figure}[htp]
\begin{center}
\scalebox{0.72}{\includegraphics*{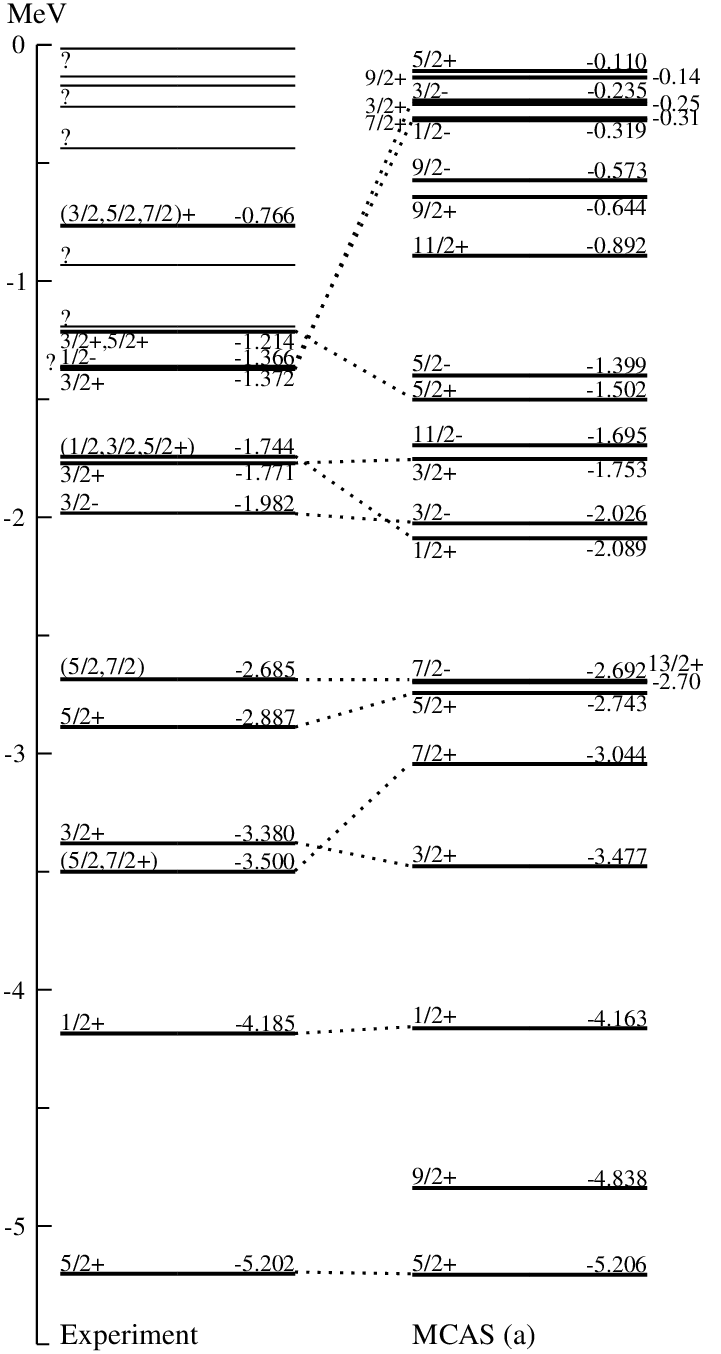}}
\end{center}
\caption{ \label{Ne23-spec} The experimental $^{23}$Ne
  spectrum~\cite{Fi07} and that calculated from MCAS evaluation of the
  $n+^{22}$Ne, with target state set (a): $0^+_1$, $2^+_1$, $4^+_1$,
  $\overline{2^+_2}$. The bar denotes the use of reduced coupling for
  channels involving this state.}
\end{figure}

The resultant MCAS spectrum of $^{23}$Ne, relative to the scattering
threshold, is shown in Fig.~\ref{Ne23-spec} for the energy range from
the ground state to the scattering threshold. A number of states in
the low-lying experimental spectrum are still not assigned with
certainty~\cite{Fi07}. With four of the fourteen states below -1 MeV
having had more than one possible $J^\pi$ suggested, and a further two
between -1.5 and -1 MeV with no conjectured $J^\pi$, this MCAS
evaluation makes an excellent match to the well-assigned states and
can make a prediction of two of the four uncertainly assigned
states. Specifically, the MCAS calculation suggests that the state at
-3.5 MeV until now denoted $(\sfrac{5}{2},\sfrac{7}{2}^+)$ is a
$\sfrac{7}{2}^+$, and the $(\sfrac{5}{2},\sfrac{7}{2})$ at -2.685 is a
$\sfrac{7}{2}^-$. For the remaining uncertainly-assigned states, MCAS
gave states with the suggested spin-parities with energies in the
proximity. We make no attempt to suggest spin-parities for states
where none has been made previously, but we do note that the density
of states above -1.5 MeV, where the number of states begins to tend
towards continuum is recreated well by MCAS.

Regarding the $\sfrac{9}{2}^+$ state found by MCAS but not seen in
experiment, it is possible that such state exists somewhere in this
regime and is as yet unobserved. This is suggested by the existence of
low-lying $\frac{9}{2}^+$ states in other mass-23 isobars. $^{23}$Na,
whose low-energy spectrum has many similarities with that of
$^{23}$Ne, has a $\sfrac{9}{2}^+$ state 2.703 MeV above the ground
state, and its mirror, $^{23}$Mg, has a state currently designated
$\sfrac{9}{2}^+,\sfrac{5}{2}^+$ at 2.714 MeV above the ground
state. Preliminary development of extensions upon this work indicate
that the $\sfrac{13}{2}^-$ at -2.696 MeV could be moved to higher
energy by employing Pauli hindrance.

\subsection{The effect of varying $\beta_L$ values in MCAS}
\label{beta2}

In Section~\ref{beta1} we illustrated an example of where it is
advantageous to extend the MCAS formalism to allow pairs of
coupling target states to have unique values of a given $\beta_L$. 
In that section, one method of selecting
the ratio of these values was outlined. Herein, we perform a
\textit{gedanken} investigation where this scaling factor spans all
values from 0 to 1.

\begin{figure}[htpb]
\begin{center}
\scalebox{0.685}{\includegraphics*{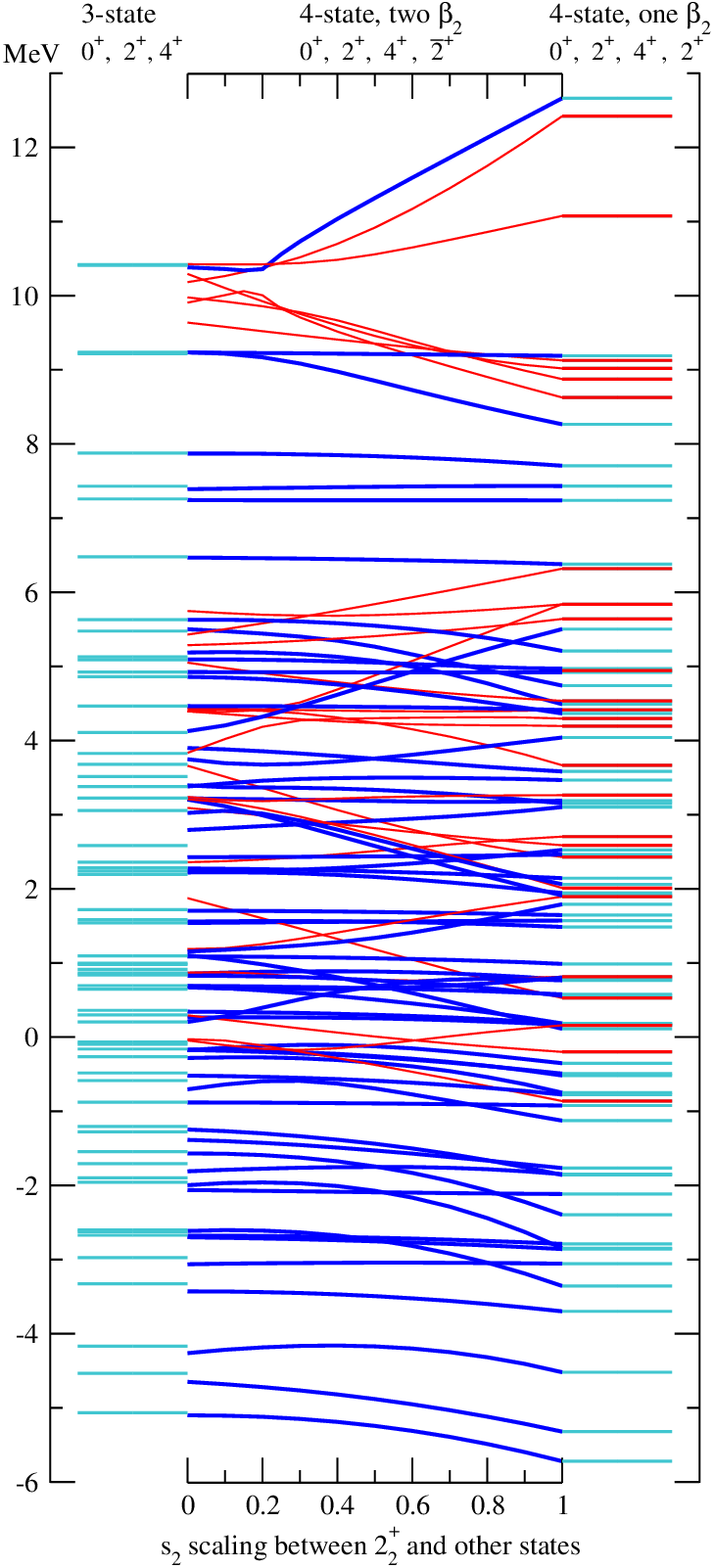}}
\end{center}
\caption{ \label{n+22Ne-2betas-ALL}(Color online.)  MCAS evaluation of
  $n+^{22}$Ne compound states with scaling of $\beta_2$ of $^{22}$Ne
  $2^+_2$ coupling. States shown by thin [red] lines are from coupling to 
  the $2^+_2$ state.}
\end{figure}

\begin{figure}[htpb]
\begin{center}
\scalebox{0.685}{\includegraphics*{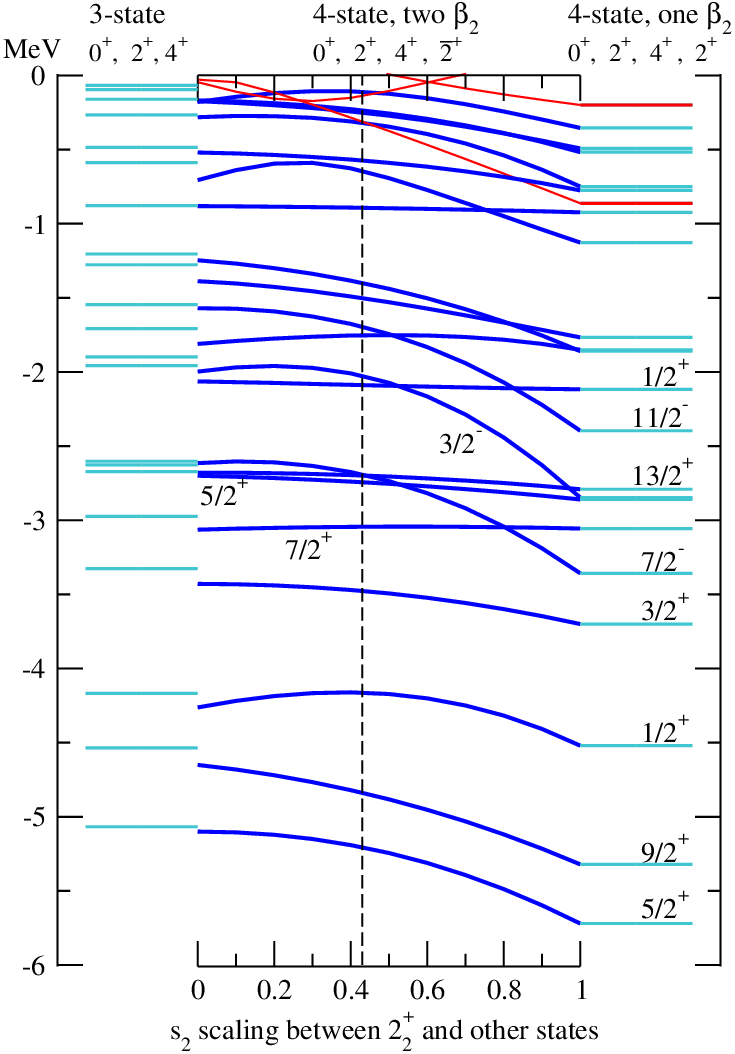}}
\end{center}
\caption{ \label{n+22Ne-2betas-LOW}(Color online.)  
  Detail of Fig.~\ref{n+22Ne-2betas-ALL}, showing subthreshold MCAS
  evaluation of the energy centroids of $n+^{22}$Ne resonances as
  $\beta_2$ of $^{22}$Ne $2^+_2$ coupling is scaled with respect to
  other couplings. The dashed line indicates value obtained from theory in
  Sec.~\ref{beta1}.}
\end{figure}

The results of diverse MCAS calculations of the spectrum of
${}^{23}$Ne to over 20 MeV excitation are given in
Figs.~\ref{n+22Ne-2betas-ALL} and \ref{n+22Ne-2betas-LOW}.  In
Fig.~\ref{n+22Ne-2betas-ALL}, both the subthreshold and resonance
parts of the spectrum are presented, while the subthreshold region is
shown in greater detail in Fig.~\ref{n+22Ne-2betas-LOW}. The first
MCAS calculation, the results of which are shown in the left most
panels of these figures identified as `3-state', used just the three
rotor-like states of ${}^{22}$Ne ($0^+_1, 2^+_1$ and $4^+_1$)
finding in all 63 states (bound and resonant) as the spectrum of
${}^{23}$Ne (with Pauli blocking included). The spectrum labelled
`4-state, one $\beta$' in these figures resulted on using additionally
the $2^+_2$ state of ${}^{22}$Ne with $\overline{\beta_2} = \beta_2$,
giving a 4-state MCAS calculation.  The spectrum that results has 89
states of ${}^{23}$Ne in the excitation energy range shown.  The
central panels show the results of 4-state MCAS calculations allowing
$\overline{\beta_2}$ to vary according to the scale variable $0 \le
s_2 \le 1$ on $\beta_2$.
The number of states for both the 3-state and 4-state calculations are
shown, by $J^\pi$, in Table~\ref{sp}.

\begin{table*}\centering
\caption{\label{sp} Number of states by $J^\pi$. Pauli blocking
  reduces the numbers of states in each case.}
\tabcolsep5pt
\begin{tabular}{l|ccccccccccccccc|c}
\hline
\hline\\[-1.9ex]
$J^\pi$ &
$\frac{1}{2}^-$ &
$\frac{1}{2}^+$ &
$\frac{3}{2}^-$ &
$\frac{3}{2}^+$ &
$\frac{5}{2}^-$ &
$\frac{5}{2}^+$ &
$\frac{7}{2}^-$ &
$\frac{7}{2}^+$ &
$\frac{9}{2}^-$ &
$\frac{9}{2}^+$ &
$\frac{11}{2}^-$ &
$\frac{11}{2}^+$ &
$\frac{13}{2}^-$ &
$\frac{13}{2}^+$ &
$\frac{15}{2}^-$&
Total\\[-1.9ex]\\
\hline\\[-1.95ex]
3-state &
4&
3&
6&
5&
7&
6&
7&
5&
6&
4&
4&
2&
2&
1&
1&
63\\
4-state&
6&
5&
9&
8&
11&
9&
10&
7&
8&
5&
5&
2&
2&
1&
1&
89\\
\hline
\hline
\end{tabular}
\end{table*}

Fig.~\ref{n+22Ne-2betas-LOW} shows the MCAS results for the
subthreshold spectrum of ${}^{23}$Ne. The spin-parities of the eleven
most bound from the 4-state evaluations are given. The effect of
changing the scale factor, $s_2$, is most noticeable, with some states
moving by as much as 2 MeV so that energy spacing and level sequence
alters. The dashed line (at $s_2 = 0.43$) indicates the spectrum found
when the ratios of $B(E2)$ values from the ground state
$\gamma$-decays of the $2^+_1$ and $2^+_2$ states define the scaling.

There are differences between the 3-state result and the result of the
4-state one when $s_2 = 0$. The disconnections are the result of two
factors. The first is that while going from right to left in the
central panels, the scaling of the $\beta_2$ is reduced from 1 to 0
but the scaling of $\beta_4 = -0.08$ is not, producing a small
difference generated by the residual $\beta_4$ coupling. The remaining
discontinuity comes from the spin-spin component of the zeroth-order
term which links channels involving different target states having the
same angular momentum, even with no deformation. This is evident in
Eq.~(\ref{Vcc-zero}) of the Appendix, wherein more details are given.

In the subthreshold region ($\lesssim -1$ MeV), the 3-state result has
equivalent states with those of the 4-state evaluations. That is not
the case for higher energies, especially as shown in
Fig.~\ref{n+22Ne-2betas-ALL} in which the above threshold resonance
centroids found from the same 3- and 4-state MCAS calculations are
displayed.  For energies $\ge -1$ MeV (see
Fig.~\ref{n+22Ne-2betas-LOW}), the 4-state model model gives 27 more
states, most of which are resonances.  The differences in resonance
centroid energies with variation of $s_2$ can be as much as 3 MeV and
the sequencing of the states alters.

Thus, in addition to what is learnt in general about the effect of the
varying the $\beta_2$ in an MCAS calculation, it is evident that the
consideration of the $^{22}$Ne $2^+_2$ has a large impact on the
ground and low-lying states of $^{23}$Ne.  Indeed, no $^{23}$Ne
spectrum from the resonances of $n+^{22}$Ne$_{0^+_1, 2^+_1, 4^+_1}$
achieved the level of agreement with data as that of the 4-state
calculation of Fig.~\ref{Ne23-spec}.  This is as would be expected,
given that the $2^+_2$ state is known to decay to the $^{22}$Ne ground
state.


\section{The effects of additional channels}
\label{otherch}

Given the result of Section~\ref{beta2}, it is instructive to examine
the influence of additional channels on results.

\subsection{Effects on the spectrum}

\begin{figure}[htp]
\begin{center}
\scalebox{0.725}{\includegraphics*{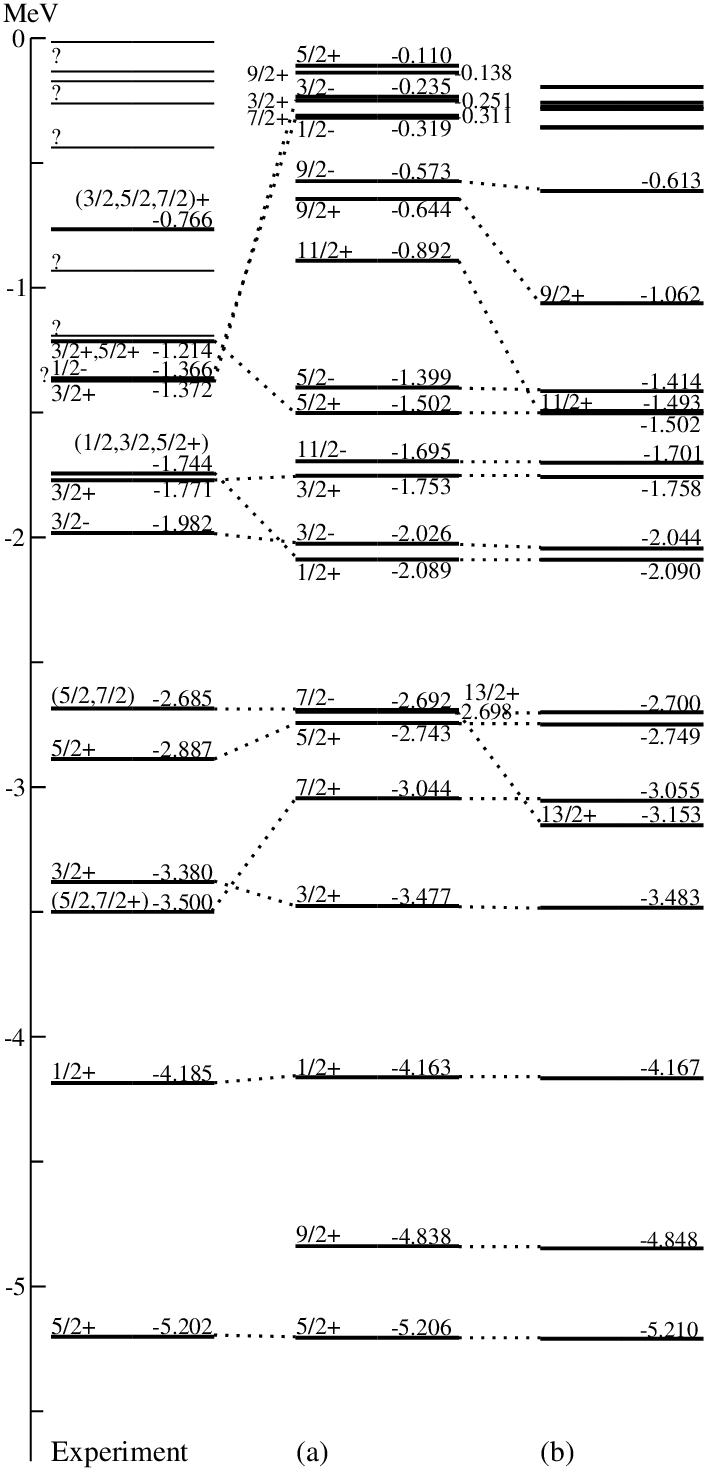}}
\end{center}
\caption{ \label{Ne23-spec-6+} Experimental data compared to MCAS
  $^{23}$Ne spectra using (a) $0^+_1$, $2^+_1$, $4^+_1$, $\overline{2^+_2}$
  [as per Fig.~\ref{Ne23-spec}]; and (b) $0^+_1$, $2^+_1$, $4^+_1$,
  $\overline{2^+_2}$, $(6)^+_1$. The bar in $\overline{2^+_2}$ denotes the use
  of reduced coupling for channels involving this state.}
\end{figure}

The impact of including the next target state in the rotation-like
band, that experimentally identified as $(6)^+$ at 6.310 MeV is now
considered. This fits well with the typical rotational band spacing of
the $0^+_1$, $2^+_1$ and $4^+_1$ states.  In Fig.~\ref{Ne23-spec-6+},
the experimental spectrum of $^{23}$Ne is compared with the MCAS
calculation result of Fig.~\ref{Ne23-spec}, denoted (a), and with that
additionally including the $6^+$ state, denoted (b). Both calculations
use the parameter set of Table~\ref{params}, which were tuned for (a),
with the Pauli blocking strengths for the $6^+$ state as per the
others. Essentially the inclusion of this 6$^+$ state affects only the
energies of the $\frac{9}{2}^+$, $\frac{11 }{2}^+$ and
$\frac{13}{2}^+$ states in this spectrum, adding to their binding.

Next we consider the influence of other states in the ${}^{22}$Ne
spectrum that may be weakly coupled to the ground state band.  We
include states deemed important in a study of the mirror system,
$p+^{22}$Mg$\rightarrow^{23}$Al~\cite{Ba11}. In Ref.~\cite{Ba11}, the
configuration mixing of the ground state of $^{23}$Al was studied
experimentally by observing the $\gamma$-rays emitted by $^{22}$Mg
after proton emission. They found the relevant components to be
$^{22}$Mg($0^+_1$)$\otimes p_{0d_{5/2}}$, with 18.5 of an observed
78.3 mb proton-emission cross section; $^{22}$Mg($2^+_1$)$\otimes
p_{1d_{5/2}}$ and $^{22}$Mg($2^+_1$)$\otimes p_{2s_{1/2}}$ with 39.3
mb; $^{22}$Mg($4^+_1$)$\otimes p_{1d_{5/2}}$ with 9.5 mb; and
$^{22}$Mg($(4)^+_2$)$\otimes p_{1d_{5/2}}$ with 10.9 of 78.3 mb
observed.  The $\gamma$-rays observed were $2^+_1 \rightarrow 0^+_1$,
$4^+_1 \rightarrow 2^+_1$, and $(4)^+_2 \rightarrow 4^+_1$, the latter
they describe as `less expected'.  Those results correspond to a
relevant spectrum of $^{22}$Mg being $0^+$ (g.s.), $2^+_1$ (1.247
MeV), $4^+_1$ (3.308 MeV), $(4)^+_2$ (5.293 MeV), the ground state,
first, second and sixth excited states. Having not observed the
relevant $\gamma$, they did not include the $2^+_2 (4.402)$ state, the
analogue of the $2^+_2$ in $^{22}$Ne, and which is known to decay via
M1+E2 to the $2^+_1$ state. It also decays by undetermined
$\gamma$-decay to the ground state and to the $4^+_1$~\cite{Fi05}.

In $^{22}$Ne, the equivalent tabulated $(4)^+_2$ state is at 5.523
MeV.  This state is denoted uncertainly as $(4)^+$, with $J=3$ also a
possibility~\cite{Fi05}, though considering the spectrum of the
mirror~\cite{Ba11,Fi05}, the $J=3$ possibility is less likely. It is
the seventh excited state, with decays to lower states
uncertain. (Ref.~\cite{Fi05} gives only that the $4^+$ $J^\pi$
assignment comes from $L=4$ in $(^6$Li$,d)$). It is shown as a dashed
line in Fig.~\ref{Ne22-spec}.

As it is unclear to what states this $(4)^+_2$ couples by
$\gamma$-emission, it is not possible to assign a $\beta_L$ value as
stringently as in Section~\ref{beta1}. Thus, the first calculation
which includes the second $4^+$ state assigns its $\beta_2$ to be that
between states in the main rotational band, so the effect of its
coupling is over- rather than under-estimated. The second calculation
assigns the same value as for the $2^+_2$ state: 43\% coupling
strength of that between states in the principal rotor
band. Fig.~\ref{Ne22-spec} graphically summarizes the states used in
the various calculations presented.

\begin{figure*}[htp]
\begin{center}
\scalebox{0.85}{\includegraphics*{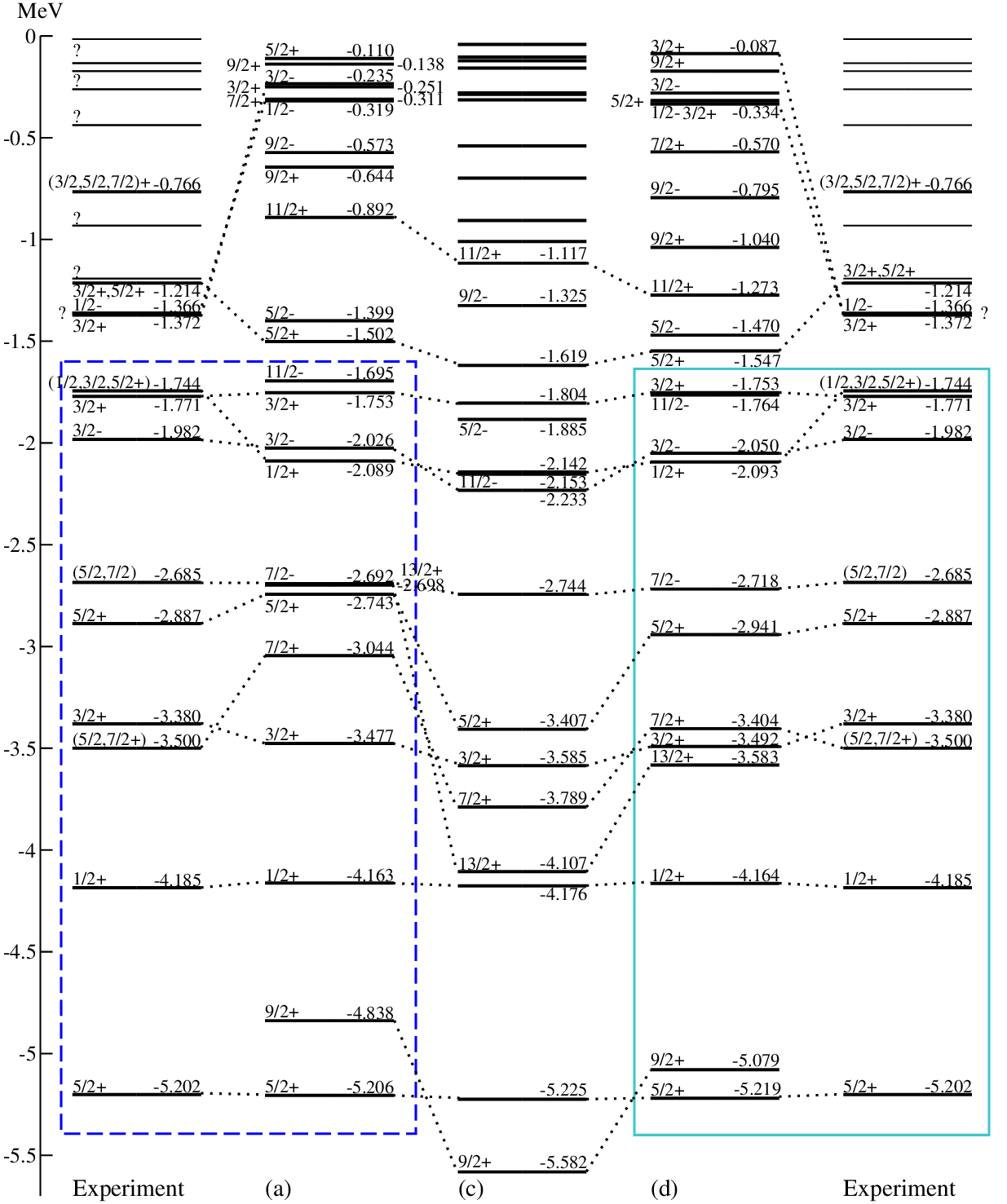}}
\end{center}
\caption{ \label{Ne23-spec-4+}(Color online.)  Experimental data
  compared to MCAS $^{23}$Ne spectra using (a) $0^+_1$, $2^+_1$,
  $4^+_1$, $\overline{2^+_2}$ [as per Fig.~\ref{Ne23-spec}]; (c) $0^+_1$,
  $2^+_1$, $4^+_1$, $\overline{2^+_2}$, $(4)^+_2$ [as per
    Ref.~\cite{Ba11}]; and (d) $0^+_1$, $2^+_1$, $4^+_1$,
  $\overline{2^+_2}$, $\overline{(4)^+_2}$. The bar denotes the use of reduced
  coupling for channels involving this state. Best result presented in
  the solid blue (online) box.}
\end{figure*}

Results of including this state in MCAS calculations are shown in
Fig.~\ref{Ne23-spec-4+}, where, it is stressed, the parameters of the
interaction potential are as listed in Table~\ref{params} for all
calculations. While the calculations including the $(4)^+_2$ do not
change the ground state energy, indicating that within this model the
ground state does not have any large component from mixing with this
state, it does show a significant influence from the $(4)^+_2$ target
state in the $\frac{7}{2}^+_1$ and $\frac{5}{2}^+_2$ compound states,
as well as those of the speculated $\frac{9}{2}^+_1$ and
$\frac{13}{2}^+_1$. A small change is also seen in the
$\frac{3}{2}^+_1$. In the case where the coupling strength is 43\% of
that within the main rotor band, denoted (d), results are improved
from the calculation (a) where the $(4)^+_2$ target state is not
included, with the $\frac{7}{2}^+_1$ and $\frac{5}{2}^+_2$ being
brought closer to observed energies.

The spectrum of $^{23}$Ne from calculation (d) and experiment,
within the energy range where this calculation is most pertinent, has been
highlighted in a solid box. This is the best result. The result of (a)
is highlighted with a dashed box.

\subsection{Effects on the cross section}

\begin{figure*}[h]
\scalebox{0.7}{\includegraphics*{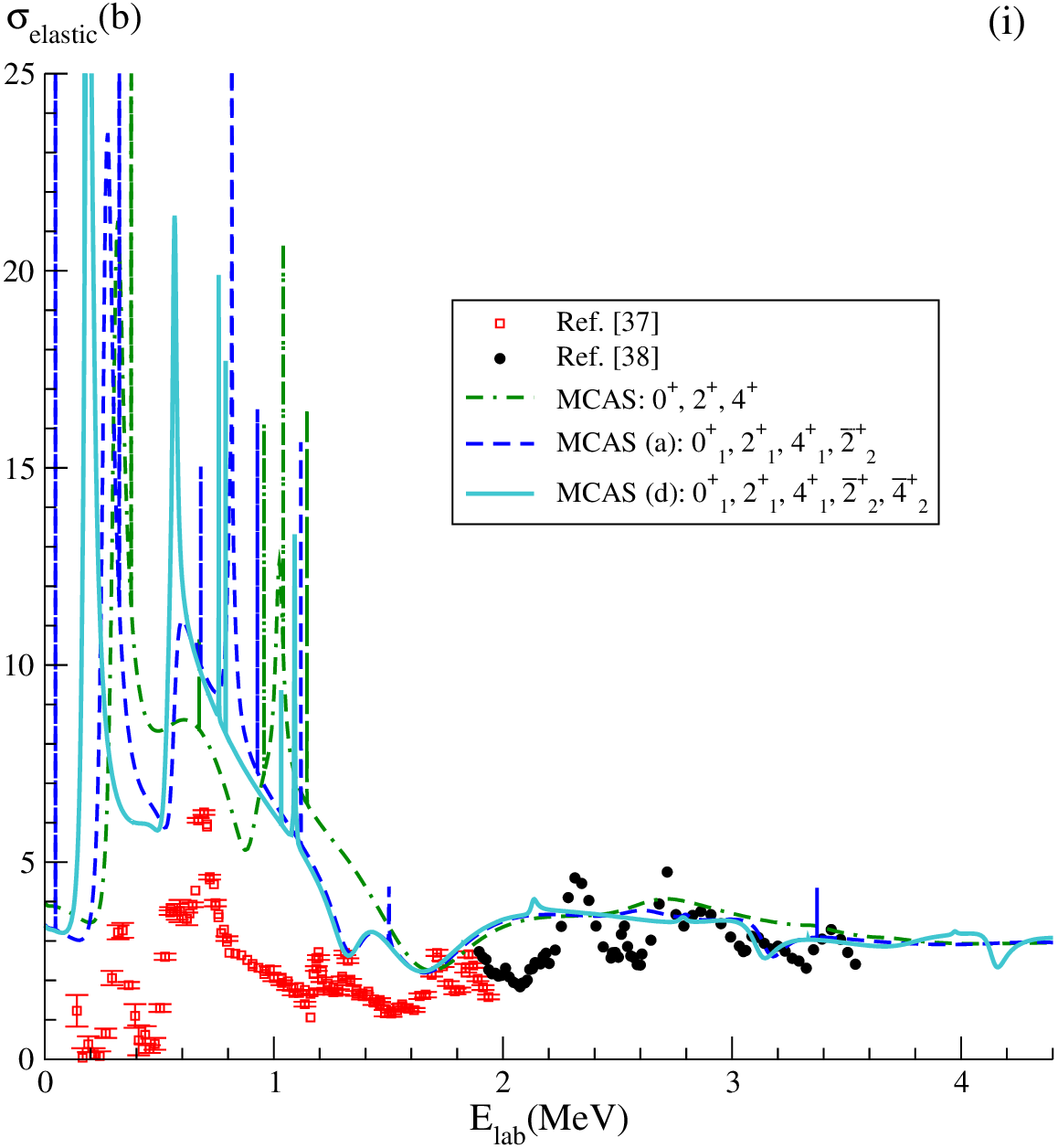}}
\scalebox{0.6}{\includegraphics*{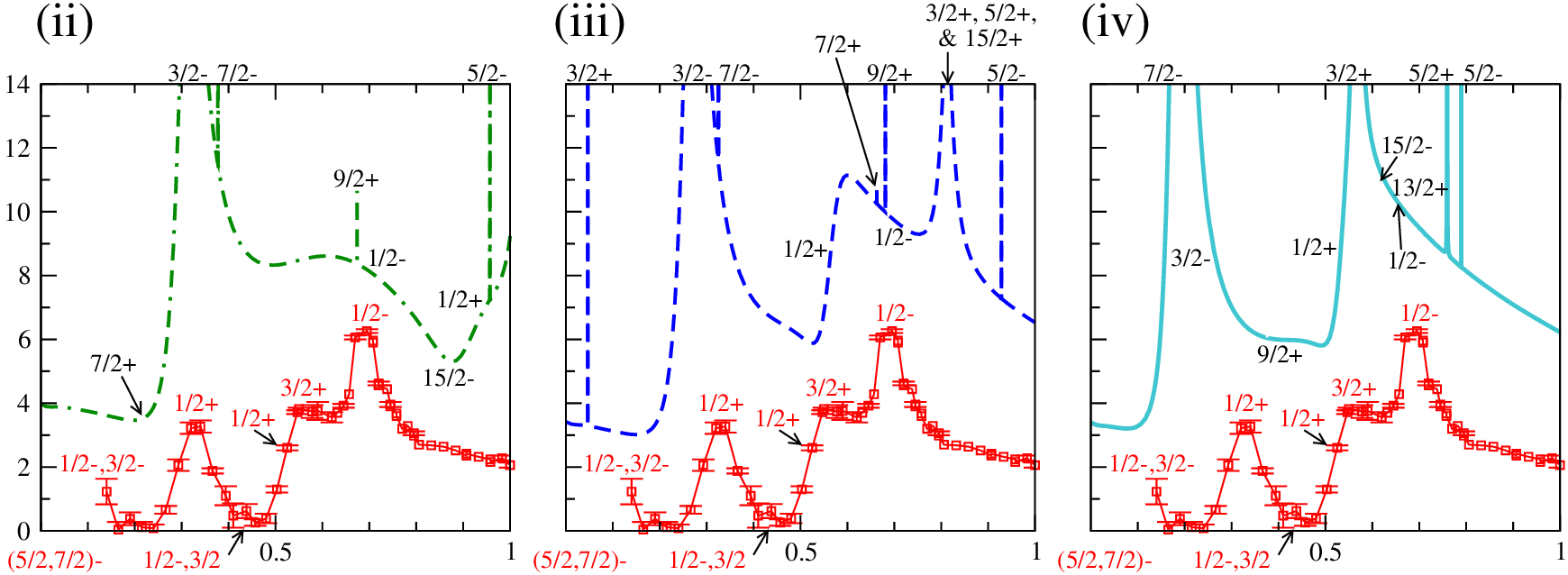}}
\caption{ \label{n+22Ne-xsect}(Color online.)  The MCAS $n+^{22}$Ne
  elastic cross section: (i) 0 - 4.4 MeV, three calculations; (ii) 0 -
  1 MeV, $0^+_1$, $2^+_1$, $4^+_1$ calculation; (iii) 0 - 1 MeV,
  $0^+_1$, $2^+_1$, $4^+_1$, $\overline{2^+_2}$ calculation
  [corresponding to (a) in Fig.~\ref{Ne23-spec}, \ref{Ne23-spec-6+}
    and \ref{Ne23-spec-4+}]; and (iv) 0 - 1 MeV, $0^+_1$, $2^+_1$,
  $4^+_1$, $\overline{2^+_2}$, $\overline{(4)^+_2}$ calculation
  [corresponding to (d) in Fig.~\ref{Ne23-spec-4+}]. (Red) squares
  from Ref.~\cite{Sa66}, black circles from Ref.~\cite{Si58}, $J^\pi$
  assignments from Ref.~\cite{Fi07}.}
\end{figure*}

The MCAS results for the $n+^{22}$Ne cross section are shown in
Fig.~\ref{n+22Ne-xsect}. It is important to note that these were not
considered during the parameter fit, and as such the results are
`predictions' of the model.

Given the density of states around this threshold, and that the $NA$
potential used has a limited energy range in which it well reproduces
results, and that this was tuned to the deeply-bound low-lying states,
the study of an MCAS elastic cross section for neutron scattering can
only be qualitative. As seen in Fig.~\ref{Ne23-spec}, this span of
energies is beyond where the current MCAS evaluations can accurately
place centroids. While only the elastic, and for incident energies
above 1.274 MeV, the inelastic scattering processes are entertained
with MCAS, the only missing possible reaction process in the energy
range examined is that of neutron capture. However, capture cross
sections for this system are negligible, with, for example,
Ref.~\cite{Be02} giving the capture background below $\sim$0.5 MeV in the
order of, at most, $100 \mu b$. Thus, the sole factor in the
over-estimation of the elastic scattering cross section MCAS result at
low energies is the limitation of the coupled-channel interaction
used.

Despite the simplicity of the chosen interaction form, MCAS has been
able to recreate some features of the observed data, which comes from
Ref.~\cite{Sa66} and \cite{Si58} for 0.14 to 1.93 MeV and 1.89 to 3.54
MeV, respectively. This is shown in Fig.~\ref{n+22Ne-xsect}, where the
marked experimental $J^\pi$ assignments are from Ref.~\cite{Fi07}. The
result of the basic $0^+_1, 2^+_1$, $4^+_1$ target state calculation
is shown by the dotted-and-dashed line, the result of calculation (a),
with the addition of the $\overline{2^+_2}$ state (the bar denoting
the weak coupling), is shown by the dashed line, and the result of
(d), with the $\overline{(4)^+_2}$, is shown by the solid line.  It
should be noted that this is a region with a high density of observed
states: as well as the seven marked $J^\pi$ values, there are four
further observed and two possibly observed resonances that have not
been assigned $J^\pi$ values between 0 and 0.6 MeV above threshold.

The upper-panel shows that, above around 1.9 MeV, MCAS recreates the
resonance background to a remarkably good degree. It also shows that
the addition of both the $2^+_2$ and $4^+_2$ states is required to
begin some representation of the resonance structure seen at this
energy.

In the lower panels, it is shown that the addition of the
$\overline{2^+_2}$ changes the shape of the calculated cross section,
and brings some resonance features into better agreement with
experiment, principally the $\sfrac{1}{2}^+$ observed on the shoulder at
0.521 MeV (lab). In the 3-state calculation we see this structure, but
at 0.905 MeV (lab). Thus, the inclusion of weakly-coupling target
states is shown to have an impact over a wide range of energies.

The addition of the $\overline{(4)^+_2}$ state brings more features into
agreement with data. In the 4-state calculation, the $\sfrac{3}{2}^+$
resonance seen in the experimental data at 0.545 MeV (lab) is located
at overly-high energy, whereas in the 5-state calculation the centroid
is brought down to the appropriate energy. The addition of the
${(4)}^+_2$ brings the calculated $\sfrac{3}{2}^-$ closer to the
uncertainly-assigned $\sfrac{1}{2}^-,\sfrac{3}{2}^-$ resonance at
0.268 MeV (lab).  Finally, in all three the cases, the
$\sfrac{1}{2}^-$ resonance observed at 0.674 MeV (lab) is detected,
though not discernible from the scattering background, and is placed
correctly in energy.

\section{Conclusions}
\label{sec-conclusion}

Historically, the multi-channel algebraic scattering formalism when
applied to a nucleon-nucleus system used a rotational collective model
to describe the selected states of the nuclear target and of the
interactions between those states and the extra nucleon.  All coupling
interactions were specified by single $\beta_L$ deformation
strengths. Most applications were of light mass systems and for a
relative small range of neutron energies in which, usually, there were
few target states deemed to belong to the main rotational band.  Use
of single values of $\beta_L$ for the selected deformations, then
sufficed to produce spectra of the compound system. For some systems
of importance, however, it is necessary to include coupling to states
outside of the main rotational band; such as cases where experiment
shows $\gamma$-decays from those extra states to ones within the
collective band.  Accordingly, MCAS has been extended to allow
coupling of different strengths between some of the set of target
states used.

This extended form of MCAS has been applied to the $n$+${}^{22}$Ne
system. The results showed that by varying the $\beta_2$ value of one
state, the $2_2^+$ (4.455 MeV), with regards to others has a marked
effect upon the evaluated spectrum of the compound, ${}^{23}$Ne.  The
value of coupling of the $2^+_2$ state with the rotor band states
($0^+_1, 2^+_1$ and $4^+_1$) was determined by using the $B(E2)$
values of the ground state $\gamma$-decays of the $2^+_1$ and $2^+_2$
states.  Treating the decays in a collective model defined the ratio
of $\overline{\beta_2} (2^+_2)$ to that of $\beta_2 (2^+_1)$. Addition
of the $(4)^+_2$ state to the target set with coupling to others
defined as being the same as the $\overline{\beta_2} (2^+_2)$,
improves the description of the spectrum as well as the
$^{22}$Ne$(n,n)^{22}$Ne elastic cross section.

The MCAS cross section recreated some resonance features observed
experimentally, and for higher energies reproduced the observed
background. The importance of coupling to target states outside of the
main rotor band was illustrated by the non-negligible changes they
make to the cross section.

The system investigated in this work is highly complex, and there are
avenues to improve the presented results in future studies. For
example, the interpretation of the Pauli principle effect, by assuming
only strictly forbidden or completely allowed shells could be
relaxed. It is known in cluster physics that the allowance of
intermediate `Pauli hindrance' is important [3]. Such consideration
may move unobserved low-lying high-spin states to higher energies,
though we cannot exclude the possibility that a $\sfrac{9}{2}^+$ state
may exist in the first few MeV of the spectrum.

\begin{acknowledgments}
PRF and LC acknowledge funds from the Dipartimento di Fisica e
Astronomia dell'Universit\`{a} di Padova and the PRIN research project
2009TWL3MX. SK acknowledges support from the National Research
Foundation of South Africa. JPS acknowledges support from the Natural
Sciences and Engineering Research Council of Canada (NSERC).
\end{acknowledgments}

\section*{Appendix}

In this appendix, further details are presented of the $NA$ scattering
potential based on a Tamura~\cite{Ta65} collective model with
rotational character for even-mass targets.

Eq.~(\ref{Mcaspot-generalL}) presented the potential, in terms of its
zeroth, first and second order components of expansion in terms of the
perturbation of the nuclear surface from spherical. It is in a form
with all terms not dependent on $L$ subsumed in equations
$v^{(0)}(r)$, $v^{(1)}(r)$ and $v^{(2)}(r)$.
\begin{align*}
V_{cc'}(r) =& V_{cc'}^{(0)}(r) + V_{cc'}^{(1)}(r) + V_{cc'}^{(2)}(r)
\nonumber\\
=& \left\{v^{(0)}(r)\right\}_{cc'}\nonumber\\  
&+\  \left\{v^{(1)}(r)\ 
\sum_{L (\ge 2)} \beta_L\ \sqrt{\frac{4\pi}{2L+1}}\  
\left[ {\bf Y}_L {\bf \cdot} {\bf Y}_L \right] 
\right\}_{cc'} 
\nonumber\\
&+\; \left\{v^{(2)}(r)\ 
\sum_{L,L' (\ge 2)} \beta_L \beta_{L'} {\sqrt{(2L+1)(2L'+1)}} \right.\nonumber\\
&\hspace{0.6cm} \left. \times \sum_{\ell} \frac{1}{2 \ell +1}\
|\langle L 0 L' 0|\ell 0 \rangle|^2\ 
[{\bf Y}_{\ell} {\bf \cdot} {\bf Y}_{\ell}]
\right\}_{cc'} .
\end{align*}

To determine the deformed channel potential, it is not simply a matter
of taking the matrix elements of the radial operators between channels
states $c$ and $c'$ and substituting them into Eq.~(\ref{www}).  The
channel potential expression involves matrix elements of the products
of two operators and so one must first make symmetric the potential
matrix form. With the zeroth order interaction, this is applicable
only to the non-diagonal term involving the operator $I\cdot s$, so
one replaces
\begin{widetext}
\begin{align}
V_{ss} w(r) \left[ I\cdot s \right]\vert_{cc'} 
&\Rightarrow \frac{1}{2} V_{ss}
\left\{ w(r) \left[ I\cdot s \right]\Vert_{c'c'}  
+ \left[ I\cdot s \right]\vert_{cc} w(r) \right\}
\nonumber\\
&\equiv
\frac{1}{2} \left[V_{ss}^{(c)} + V_{ss}^{(c')}\right]
 w(r) \left[ I\cdot s \right]\vert_{cc'} .
\end{align}
Thus, the zero order term in Eq.~(\ref{Mcaspot-generalL}) is
\begin{equation}
\left\{ V^{(0)}(r) \right\}_{c c^\prime} = \left\{ 
\left[ V_0^{(c)} + V_{ll}^{(c)} l(l+1) \right] w(r)
+ W_{ls}^{(c)} \frac{1}{r} \pdif{w(r)}{r} {\{\bf l \cdot s}\}
\right\} \delta_{c c^\prime}
+ \frac{1}{2} \left[ V_{ss}^{(c)} + V_{ss}^{(c^\prime)} \right]
\  {\{\bf I \cdot s}\}_{cc^\prime}\  w(r) \ .
\label{Vcc-zero}
\end{equation}
\end{widetext}
The $\delta_{cc'}$ is added to stress that the included terms
contribute only on the diagonal. (This potential accounts for some of
the discontinuities between the 3- and 4-state calculations in
Fig.~\ref{n+22Ne-2betas-ALL} and \ref{n+22Ne-2betas-LOW} when
$s_2=0$. While the first term does not contribute, being diagonal in
channels, the dependence of the second term on $\{{\bf I \cdot
s}\}_{cc^\prime}$ has the consequence that, even with no deformation,
channels of the same spin-parity $I^\pi$ are coupled. Thus, the
$2^+_2$ state is linked to the $2^+_1$.)

In the first and second order terms, the other two components also need be
taken with symmetrised operators, whence
\begin{widetext}
\begin{align}
\left\{ V^{(1)}(r)\right\}_{cc^\prime} = 
&-R_0 \left\{ \pdif{w(r)}{r} \
\ \frac{1}{2} \left[ V_0^{(c)} + V_0^{(c^\prime)} 
 +\; V_{ll}^{(c)} \left\{{\bf l\cdot l}\right\}_{cc} 
+ V_{ll}^{(c^\prime)} \left\{{\bf l\cdot l}\right\}_{c^\prime c^\prime}\right] \right.
\nonumber\\
&\hspace{3cm} \left.- \frac{1}{2}\ R_0 \frac{1}{r}\ 
\frac{\partial^2 w(r)}{\partial r^2}  
\Bigl( W_{ls}^{(c)} \left\{{\bf l\cdot s}\right\}_{cc}
+ W_{ls}^{(c^\prime)} \left\{{\bf l\cdot s}\right\}_{c^\prime c^\prime}\Bigr)
\right\}
\left[ \sqrt{\frac{4\pi}{2L+1}} \beta_L
\left[ {\bf Y}_L {\bf \cdot} {\bf Y}_L \right]
\right]_{cc^\prime}
\nonumber\\
&- \frac{1}{2}\ R_0 \pdif{w(r)}{r} \sum_{c^{\prime \prime}} 
\left\{ V_{ss}^{(c^\prime)}
\left[
\sqrt{\frac{4\pi}{2L+1}} \beta_L
\left[ {\bf Y}_L {\bf \cdot} {\bf Y}_L \right]
\right]_{cc^{\prime\prime}}
\left[{\bf I \cdot s}\right]_{c^{\prime \prime}c^\prime}
\right.
\nonumber\\
&\hspace{8cm}\left.
+ V_{ss}^{(c)}
\left[{\bf I \cdot s}\right]_{c c^{\prime \prime}}
\left[
\sqrt{\frac{4\pi}{2L+1}} \beta_L
\left[ {\bf Y}_L {\bf \cdot} {\bf Y}_L \right]
\right]_{c^{\prime \prime}c^\prime}
\right\} ,
\label{Vcc-one}
\end{align}
and
\begin{align}
\left\{ V^{(2)}(r)\right\}_{cc^\prime} =& \: R_0^2 \left\{
\frac{1}{4} \pdif{^2w(r)}{r^2} \left[ 
V_0^{(c)} + V_0^{(c^\prime)} 
+ V_{ll}^{(c)} \left\{{\bf l\cdot l}\right\}_{cc}
+ V_{ll}^{(c^\prime)} \left\{{\bf l\cdot l}\right\}_{c^\prime c^\prime}
\right] \right.
\nonumber\\
&+ \frac{R_0^2}{4}\ \left. \frac{1}{r}\ 
\frac{\partial^3 w(r)}{\partial r^3}  
\Bigl(W_{ls}^{(c)}\ \left\{{\bf l\cdot s}\right\}_{cc}
+ W_{ls}^{(c^\prime)}\ \left\{{\bf l\cdot s}\right\}_{c^\prime c^\prime} 
\Bigr) \right\}
\nonumber\\
&\hspace{4cm}\times
\left[ 
\sum_{L L'} \beta_L \beta_{L'} {\sqrt{(2L+1)(2L'+1)}} \sum^{2L}_{\ell\ {\rm even}} 
\frac{1}{(2\ell+1)}
|\langle L 0 L' 0|\ell 0 \rangle|^2\
\left[ {\bf Y}_\ell  {\bf \cdot} {\bf Y}_\ell \right]
\right]_{cc^\prime}
\nonumber\\
&\hspace{-1cm}+ \frac{R_0^2}{4}\ \pdif{^2w(r)}{r^2} 
\sum_{c^{\prime \prime}} \left\{ V_{ss}^{(c)}\ 
\left\{{\bf I\cdot s}\right\}_{cc^{\prime \prime}} 
\left[
\sum_{L L'} \beta_L \beta_{L'} {\sqrt{(2L+1)(2L'+1)}} \sum^{2L}_{\ell\ {\rm even}} 
\frac{1}{(2\ell+1)}
|\langle L 0 L' 0|\ell 0 \rangle|^2\
\left[ {\bf Y}_\ell  {\bf \cdot} {\bf Y}_\ell \right]
\right]_{c^{\prime \prime}c^\prime}
\right.\nonumber\\
&\left.
+ V_{ss}^{(c^\prime)}\
\left[
\sum_{L L'} \beta_L \beta_{L'} {\sqrt{(2L+1)(2L'+1)}} \sum^{2L}_{\ell\ {\rm even}} 
\frac{1}{(2\ell+1)}
|\langle L 0 L' 0|\ell 0 \rangle|^2\
\left[ {\bf Y}_\ell  {\bf \cdot} {\bf Y}_\ell \right]
\right]_{c c^{\prime \prime}}
\left\{{\bf I\cdot s}\right\}_{c^{\prime \prime} c^\prime} 
\right\}\ .
\label{Vcc-two}
\end{align}
\end{widetext}
The matrix elements of the operators 
${\bf l \cdot l}$, ${\bf I \cdot s }$, and ${\bf l \cdot s}$
are 
\begin{equation}
\langle {\bf l \cdot l} \rangle =
\langle l' j' I' J | {\bf l \cdot l}
| l j I J \rangle
= \delta_{ll'}\delta_{jj'}\delta_{II'} l(l + 1)\ ,
\end{equation}
and
\begin{equation}
\langle {\bf s \cdot l} \rangle =
\delta_{ll'}\delta_{jj'}\delta_{II'} \times
\left\{
\begin{array}{ccc}
\frac{l}{2} & {\rm , if} & j=l+\frac{1}{2}\\
-\frac{l+1}{2} & {\rm , if} & j=l-\frac{1}{2}\\
\end{array}
\right.
.
\end{equation}
The spin-spin matrix element is more complicated~\cite{Va88}, namely:

\begin{widetext}
\begin{eqnarray}
\langle {\bf s} {\bf \cdot} {\bf I} \rangle
&=& (-)^{(j+j'+J)} \left\{
\begin{array}{ccc}
j' & j & 1\\
I & I' & J\\
\end{array}
\right\}
\left\langle I' \left\| {\bf I} \right\| I \right\rangle
\left\langle s \left\| {\bf s} \right\| s \right\rangle
\nonumber \\
&=& \delta_{II'} \delta_{ll'}  (-)^{\left(1/2 + j - j'+ I + J + l\right)}
\sqrt{(2j+1) (2j'+1) (2I+1)}
\sqrt{\frac{3}{2} I(I+1)}
\left\{
\begin{array}{ccc}
j' & j & 1\\
I & I & J\\
\end{array}
\right\}
\left\{
\begin{array}{ccc}
\frac{1}{2} & l & j\\
j' & 1 & \frac{1}{2}\\
\end{array}
\right\} \ .
\end{eqnarray}
\end{widetext}
The operator is diagonal in I and $l$, and zero if either I or I' is zero.

Finally, the
matrix elements of the scalar product of two rank L spherical harmonics
are needed. They are 
\begin{widetext}
\begin{eqnarray}
\langle {\bf Y}_L {\bf \cdot} {\bf Y}_L\rangle &=&
\left\langle {l'j'I'J} \left| {\bf Y}_L(\hat r) {\bf \cdot}
{\bf Y}_L(\hat \Upsilon)
\right| ljIJ \right\rangle \nonumber\\
&=& (-)^{\left( j + I' + J \right)}
\left\{
\begin{array}{ccc}
j' & j & L\\
I & I' & J\\
\end{array}
\right\}
\left\langle \left(l'{\sfrac{1}{2}}\right)j' \right\| {\bf Y}_L(\hat r) \left\|
\left(l{\sfrac{1}{2}}\right)j \right\rangle
\left\langle I' \left\| {\bf Y}_L(\hat \Upsilon)
\right\| I \right\rangle \nonumber \\
&=& (-)^{\left(j+I'+l'-{\frac{1}{2}}\right)} \sqrt{(2j+1) (2j'+1) 
(2I+1) (2l+1)}
\nonumber \\
&&\hspace{4.5cm}
\times \frac {1}{4\pi} (2L+1)
\left\langle I 0 L 0 \vert I^\prime 0 \right\rangle
\left\langle l 0 L 0 \vert l^\prime 0 \right\rangle
\left\{
\begin{array}{ccc}
j' & j & L\\
I & I' & J\\
\end{array}
\right\}
\left\{
\begin{array}{ccc}
l & \frac{1}{2} & j\\
j' & L & l'\\
\end{array}
\right\} \ ,
\label{tensor}
\end{eqnarray}
which, on using the identity
\begin{equation}
\left\{
\begin{array}{ccc}
l & \frac{1}{2} & j\\
j' & L & l'\\
\end{array}
\right\}
\left\langle l 0 L 0 \vert l^\prime 0 \right\rangle
= (-)^{\left( l+j'+\frac{1}{2}\right)} \frac {1}{\sqrt{(2l+1)(2j'+1)}}
\left\langle j {\sfrac{1}{2}} L 0 \right|\left. j^\prime {\sfrac{1}{2}} 
\right\rangle \ ,
\end{equation}
reduce to
\begin{align}
&\left\langle l'j'I'J \left| {\bf Y}_L({\hat r}) {\bf \cdot}
{\bf Y}_L({\hat \Upsilon}) \right|
ljIJ \right\rangle =\ (-)^{\left(J-{\frac{1}{2}+I'}\right)} \frac {1}{4\pi}
\sqrt{(2I+1) (2j+1) (2j'+1) (2L+1)}\nonumber\\
&\hspace{7cm}\times \frac{1}{2}
\left[1 + (-)^{l + l' + L} \right]\ 
\left\langle I 0 L 0 \vert I^\prime 0 \right\rangle
\left\langle j {-\sfrac{1}{2}} j^\prime {\sfrac{1}{2}} \left| \right. L 
0  \right\rangle
\left\{
\begin{array}{ccc}
j' & j & L\\
I & I' & J\\
\end{array}
\right\} \ .
\end{align}
\end{widetext}

\bibliography{Fr14}

\end{document}